# Avalanche effect and gain saturation in high harmonic generation

Carles Serrat,[1,*] David Roca,[1] Josep M. Budesca,[1] Jozsef Seres,[2] Enikoe Seres,[2] Bastian Aurand,[3] Andreas Hoffmann,[4] Shinichi Namba,[5] Thomas Kuehl,[6,7] and Christian Spielmann,[4,7]

[1] Universitat Politècnica de Catalunya, Departament de Física i Enginyeria Nuclear, Colom 11, 08222 Terrassa, Spain.

[2] Institute of Atomic and Subatomic Physics, Vienna University of Technology, Stadionalle 2, 1020 Vienna, Austria

[3] Institut für Laser- und Plasmaphysik, Universität Düsseldorf, Universitätsstr. 1, 40225 Düsseldorf, Germany

[4] Institute of Optics and Quantum Electronics, Abbe Center of Photonics, Friedrich Schiller University, Max Wien Platz 1, 07743 Jena, Germany

[5] Graduate School of Engineering, Hiroshima University, 1-4-1 Kagamiyama, Higashi-Hiroshima, Hiroshima 739-8527, Japan

[6] GSI Helmholtz Centre for Heavy Ion Research, Planckstrasse 1, 64291 Darmstadt, Germany

[7] Helmholtz Institute Jena, Fröbelstieg 3, 07743 Jena, Germany

Corresponding author: Carles Serrat, e-mail: carles.serrat-jurado@upc.edu

Optical amplifiers in all ranges of the electromagnetic spectrum exhibit two essential characteristics: i) the input signal during the propagation in the medium is multiplied by the avalanche effect of the stimulated emission to produce exponential growth and ii) the amplification saturates at increasing input signal. We demonstrate that the strong-field theory in the frame of high harmonic generation fully supports the appearance of both the avalanche and saturation effects in the amplification of extreme ultraviolet attosecond pulse trains. We confirm that the amplification takes place only if the seed pulses are perfectly synchronized with the driving strong field in the amplifier. We performed an experimental study and subsequent model calculation on He gas driven by intense 30-fs-long laser pulses, which was seeded with an attosecond pulse train at 110 eV generated in a separated Ne gas jet. The comparison of the performed calculations with the measurements clearly demonstrates that the pumped He gas medium acted as an amplifier of the extreme ultraviolet pulse train.

## I. INTRODUCTION

High harmonic generation (HHG) converts intense, short laser pulses to their harmonics and generates coherent radiation in the extreme ultraviolet (XUV) and soft X-ray spectral range. HHG is very flexible and able to fulfil the demand of different applications, viz. it can generate very short pulses with durations even of attoseconds [1,2], or very high harmonics with energies of few keV [3-6] for time resolved spectroscopy [7,8].

In order to tailor the spectral shape or the temporal profile of the HHG pulses and to improve the pulse energy according to the different demands, HHG is extensively studied. A very promising method consists in the illumination of the gas used for HHG with a vacuum ultraviolet (VUV) or XUV pulse together with a high-intensity infrared (IR) or near-infrared (NIR), short laser pulse. Such VUV/XUV pulse, generated by an independent source [9-11] or within a gas mixture [12], was used to enhance or synchronize the ionization of the gas atoms, and strong enhancement of HHG was reported.

Nonlinear parametric processes in HHG involving the short NIR laser pulse and the XUV pulse were reported in Ref. [13]. Such nonlinear interaction, as X-ray parametric amplification (XPA), can cause amplification of the XUV pulse in the gas medium. Both nonlinear enhanced ionization and stimulated amplification were theoretically studied, namely, a XUV seed pulse was shown to produce new harmonic lines [14], and it was shown to be amplified by backward scattering [15] or forward scattering [16-20]. Parametric amplification processes have been recently measured and described also by perturbative high-order parametric interaction [21].

Here we show that the experimentally found amplification of coherent attosecond XUV pulses in He gas is fully supported by numerical simulations based on the quantum-mechanical description of HHG using the strong field approximation (SFA) [22]. Indeed, in Ref. [18] it was revealed that the amplification of coherent XUV attosecond pulses can be obtained by synchronizing a weak XUV pulse with the strong IR pulse. This theoretical prediction was soon corroborated by the experiments in Ref. [23], which measured XUV attosecond pulse amplification in He gas at around 110 eV photon energies. In the present work we extend the research in Refs. [18] and [23] and show large agreement between theoretical and experimental observations concerning saturation of the amplification by increasing seed intensity and amplification by avalanche of a XUV attosecond pulse train in an He gas amplifier at the 110 eV region. The numerical simulations show that plasma dispersion in the amplifying medium and the ionization potential of the gas [19] are key factors to produce XUV amplification in a specific spectral region.

## II. EXPERIMENTAL SETUP

In order to study parametric amplification of an XUV attosecond pulse train, the experiments were performed using a Ti:sapphire laser system delivering 30 fs pulses with a central wavelength of 800 nm and 30 mJ of energy at 10 Hz repetition rate (see Fig. 1). The pulses were loosely focused to obtain an intensity of ~$10^{15}$ W/cm². The HHG source consisted of two independent gas jets: the harmonics generated in the first gas jet in the form of an attosecond pulse train served as the XUV seed, and the second gas jet served as amplifier. The beam profiles and the spectra of the harmonics generated in the second gas jet were measured at around 110 eV. The laser light and the low order harmonics were filtered out by thin metal foils of 200-nm-thick Zr and 200-nm-thick Ti for beam profile measurements, and two pieces of 300-nm-thick Zr foil were used for spectral measurements. The beam profiles described in more detail in section 4A and the spectra shown in section 4B were taken with 10 s and 50 s integration times, respectively. The seed jet was filled with neon in order to produce a suitable intense seed beam for saturating the amplifier when it was necessary. The intensity of the seed beam was controlled by adjusting the Ne gas pressure. The gas medium in the amplifier jet was helium also with adjustable pressure. During the experiments, the backing pressure of Ne and He was adjusted up to 1.2 bar and 5 bar, respectively while the gas pressure in the interaction volume was about 2-4 % of the backing pressure (see Method section of [23]). A more detailed description of the setup; the calibration of the measured XUV fluences can be found in the Method section of Ref. [23].

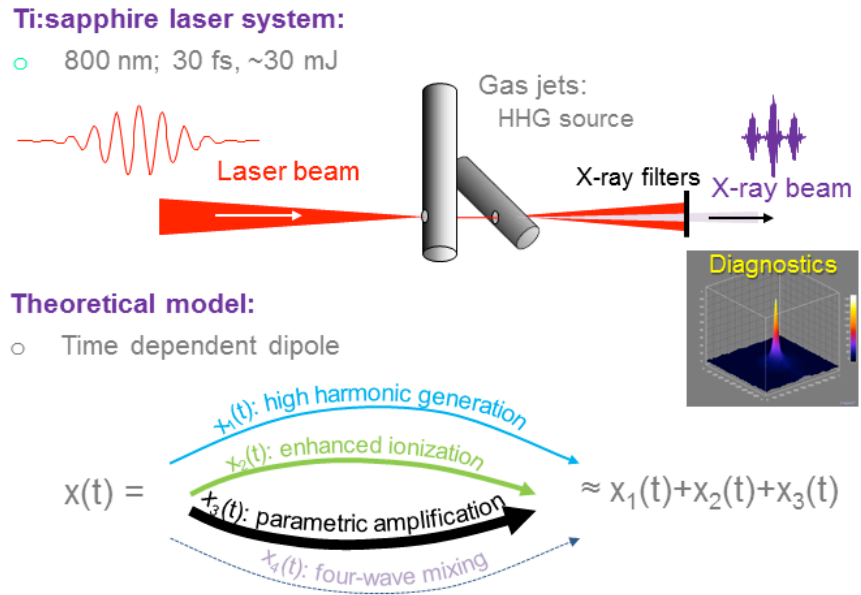

FIG. 1. (Color online). Experimental setup and the theoretical method. Both the experiments and the theory are based on an assembly of two jets for achieving and describing amplification of attosecond pulse trains. By decomposing the dipole matrix elements of the time-dependent dipole moment four different scattering processes can be identified, as indicated, of which $x_4$ is much smaller than the others and can be neglected (see text for details).

## III. THEORETICAL MODEL AND SINGLE-ATOM SIMULATIONS

Our theoretical model is based on an extension of the single-atom response calculated by solving the Schrödinger equation in the SFA in the nonadiabatic form, so that the full electric field of the laser pulse is used to calculate the nonlinear dipole moment [22]. In order to describe high-order harmonic generation produced in the tunnel and over-the-barrier ionization regimes, i.e. for $U_p > I_p >> \omega$, with $U_p$ being the ponderomotive energy of the electron in a laser field of frequency $\omega$ and $I_p$ being the atomic ionization potential, the SFA approach makes the following assumptions: i) all bound states in the atom except the ground state are neglected; ii) the depletion of the ground state is neglected by considering $U_p < U_{sat}$, with $U_{sat}$ being the saturation level; iii) in the continuum the electron is assumed free from the influence of the atomic potential; and iv) the less singular part of the continuum-continuum transitions between states of different energies in the dipole matrix element is neglected, while the motion of the free electron in the laser field is treated exactly. These assumptions have been shown to provide good qualitative agreement with the numerical solution of the Schrödinger equation for high-harmonics produced at photon energies significantly larger than the atomic ionization potential $I_p$. The SFA theory thus considers the interaction of an intense low-frequency field with the medium for the approximate description of the high-harmonics generation processes but it makes no explicit approximation on the frequency of the laser field, and also accounts for most important quantum effects - such as quantum diffusion of wave packets and quantum interferences.

In the extension of the theory that follows we consider the intense low-frequency IR field generating the high-harmonics perturbed by a weak XUV electric field which ionization probability is negligible compared to the ionization produced by the IR field. As it will be shown, this small perturbation allows us to describe the contribution of different nonlinear scattering processes induced by the weak XUV field, which are well understood and accurately described by our extended theory, since they are not significantly affected by the assumptions made in the SFA theory for the description of the purely high-harmonics generation processes.

Following [22], after solving the Schrödinger equation and by considering the stationary values of the classical action in the frame of the saddle-point approximation for the canonical momentum, the time-dependent dipole moment can be written in the form of Eq. (13) in Ref. [22]:

$$x(t) = i\int_0^t dt' \left(\frac{\pi}{\varepsilon + i(t-t')/2}\right)^{3/2} d^*[p_{st}(t,t') - A(t)] \times e^{-iS_{st}(t,t')} d[p_{st}(t,t') - A(t')]E(t') + c.c. \quad (1)$$

where

$$p_{st}(t,t') = \frac{1}{t-t'}\int_{t'}^{t} dt''\, A(t''), \quad (2)$$

is the stationary value of the canonical momentum and

$$S_{st}(t,t') = I_p(t-t') - \frac{1}{2}p_{st}^2(t,t')(t-t') + \frac{1}{2}\int_{t'}^{t} dt''\, A^2(t''). \quad (3)$$

is the stationary value of the action. $A(t) = -\int_0^t dt' E(t')$ is the vector potential of the laser field, which is considered linearly polarized in the x-direction, $I_p$ is the atomic ionization potential and $\varepsilon$ is an infinitesimal constant. We will consider the case of hydrogen-like atoms, for which the dipole matrix element for transitions to and from the continuum with momentum $k$ can be approximated [22] as

$$d(k) = i\frac{2^{7/2}(2I_p)^{5/4}}{\pi}\frac{k}{(k^2+2I_p)^3}. \tag{4}$$

The driving laser field in our study is composed by a strong femtosecond IR pulse and an attosecond high-frequency weak XUV single pulse or train of pulses $E(t) = E_{IR}(t) + E_{XUV}(t)$, so $A(t) = A_{IR}(t) + A_{XUV}(t)$ and we can write the time-dependent dipole moment as $x(t) = x_{IR}(t) + x_{XUV}(t)$, with

$$\begin{aligned}x_{IR,XUV}(t) = i\int_0^t dt'&\left(\frac{\pi}{\varepsilon+i(t-t')/2}\right)^{3/2} d^*[p_{st}(t,t')-A(t)] \\ \times e^{-iS_{st}(t,t')}&d[p_{st}(t,t')-A(t')]E_{IR,XUV}(t') + c.c.\end{aligned} \tag{5}$$

The contribution from $x_{XUV}(t)$ to the total time-dependent dipole moment $x(t)$ can be neglected if the amplitude of the XUV field is small, as it is the case considered in our study for the single-atom interaction or for small pressures and/or small propagation distances. Therefore, in the case that the XUV field is weak, we can approximate

$$\begin{aligned}x(t) \approx x_{IR}(t) = i\int_0^t dt'&\left(\frac{\pi}{\varepsilon+i(t-t')/2}\right)^{3/2} d^*[p_{st}(t,t')-A(t)] \\ \times e^{-iS_{st}(t,t')}&d[p_{st}(t,t')-A(t')]E_{IR}(t') + c.c.\end{aligned} \tag{6}$$

In the calculation of the dipole matrix elements [Eq. (4)] however, the XUV field cannot be neglected since, as it will be shown below in detail, it results in nonlinear parametric processes that amplify the XUV signal giving an essential contribution to the generated harmonic signal.

**A. Decomposition of the dipole moment**

Let us further study the particular processes driving the time-dependent dipole moment [19]. By including $k = p_{st} - A_{IR} - A_{XUV}$ into Eq. (4), we can write Eq. (6) as

$$x_{IR}(t) \approx i\int_0^t dt' \left(\frac{2^{14}(2I_p)^5}{\pi(\varepsilon + i(t-t')/2)^3}\right)^{1/2} e^{-iS_{IR}^{st}(t,t')} E_{IR}(t') \times$$
$$\frac{d_1(t,t') + d_2(t,t') + d_3(t,t') + d_4(t,t')}{([p_{IR}^{st}(t,t') - A_{IR}(t)]^2 + 2I_p)^3 ([p_{IR}^{st}(t,t') - A_{IR}(t')]^2 + 2I_p)^3} + c.c.,$$  (7)

where

$$d_1(t,t') = [p_{IR}^{st}(t,t') - A_{IR}(t)][p_{IR}^{st}(t,t') - A_{IR}(t')],$$  (8)

$$d_2(t,t') = -[p_{IR}^{st}(t,t') - A_{IR}(t)]A_{XUV}(t'),$$  (9)

$$d_3(t,t') = -[p_{IR}^{st}(t,t') - A_{IR}(t')]A_{XUV}(t),$$  (10)

$$d_4(t,t') = A_{XUV}(t)A_{XUV}(t').$$  (11)

In the previous expressions we have assumed

$$p_{st}(t,t') \approx p_{IR}^{st}(t,t') = \frac{1}{t-t'}\int_{t'}^t dt'' A_{IR}(t''),$$  (12)

$$S_{st}(t,t') \approx S_{IR}^{st}(t,t') = I_p(t-t') - \frac{1}{2}(p_{IR}^{st})^2(t,t')(t-t') + \frac{1}{2}\int_{t'}^{t} dt'' A_{IR}^2(t''), \tag{13}$$

and in the denominator of the dipole matrix elements

$$A = A_{IR} + A_{XUV} \approx A_{IR} \tag{14}$$

has been taken. The approximations in Eqs. (6), (7), and (12) - (14) are accurate for single-atom interactions with the parameter values considered in our simulations and they allow us to determine with precision the contributions of the IR and XUV fields in the theory. To calculate the spectra in our study, however, we proceed as follows: For the single atom interaction, numerical integration of Eq. (1) has been performed to compute the time-dependent dipole moment $x(t)$. The Fourier transform of the acceleration of the dipole $d^2x(t)/dt^2$ gives the field spectrum $x(\omega)$ from which the power spectrum $P(\omega) = |x(\omega)|^2$ is calculated. We have considered ground-state depletion by using the tunnel ionization rate in the ADK theory [24].

The decomposition of the dipole matrix element written in Eq. (12) therefore provides four integrals $x_1$ - $x_4$ corresponding to the $d_1 - d_4$ terms in Eqs. (13) - (16), respectively, that we can compute separately. Fig. 2 shows the contribution of these integrals to $x(t)$ for a single Gaussian XUV pulse [Figs. 2(a, c, e)] and for a XUV pulse train [Figs. 2(b, d, f)], both centered at $\simeq$ 113 eV, in a single atom calculation. We observe that $x_1$ [blue solid line in Fig. 2(c) and (d)] accounts for the regular high-harmonic generation processes, i.e. the spectrum that would be obtained in the absence of the $E_{XUV}(t)$ field.

The contribution from $x_2$ computes the probability of the release of the electron from the atom by both the presence of laser field $E_{IR}(t')$ and the XUV attosecond pulse field through its vector potential $A_{XUV}(t')$ at the time $t'$, the propagation to time $t$ by the semiclassical action $S_{IR}^{st}(t,t')$ and the recombination at time $t$. Some of these processes concerning photon energies near the ionization potential of the gas medium have been extensively studied in the last years [9-12]. In these experiments, the first HHG source was optimized to produce intense low-order or bellow threshold harmonics to affect efficiently the ionization process and to enhance HHG in the XUV.

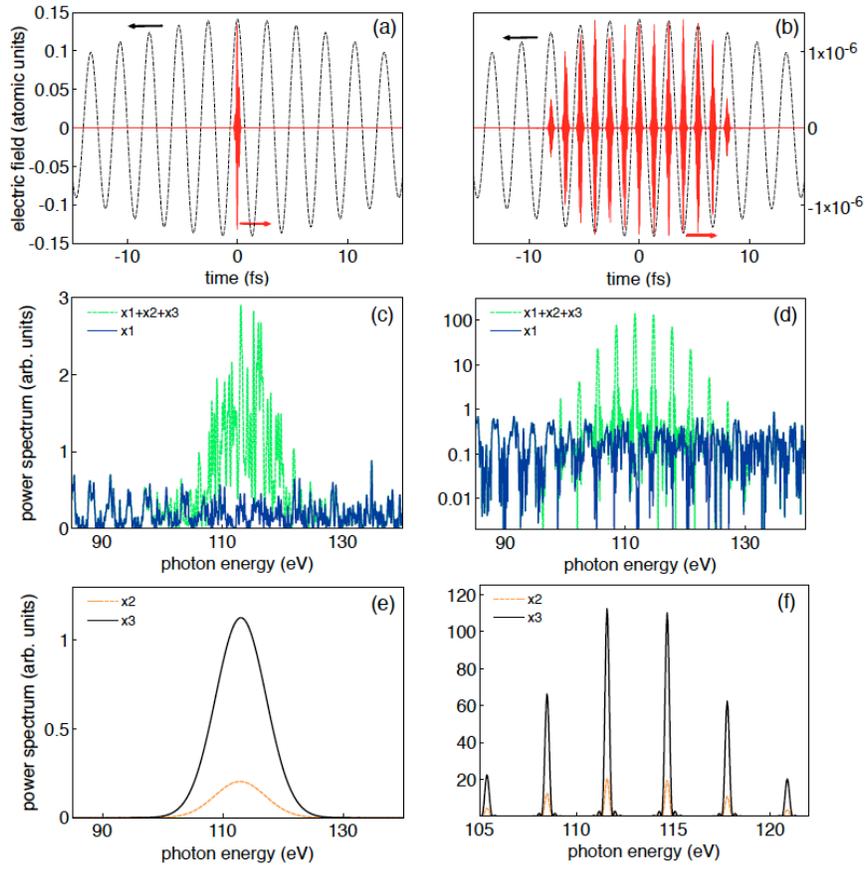

FIG. 2. (Color online). Single atom calculations. The case of a single attosecond XUV pulse [(a), (c) and (e)] and a train of XUV pulses [(b), (d) and (f)] in He ($I_p$=24.587 eV) was calculated. In (a) and (b) the IR and XUV fields are shown. The IR pulse is an 800 nm, $7\times10^{14}$ W/cm$^2$ peak intensity pulse of 26 fs (FWHM) Gaussian temporal profile [black dotted lines in (a) and (b)].

The XUV field consists of Gaussian 200 as pulses, with peak intensity of $7\times10^4$ W/cm$^2$ (i.e. $10^{-10}$ times the IR peak intensity). In (b) the pulse train has a super Gaussian envelope of 15 fs. The IR carrier envelope phase (CEP) is perfectly synchronized with the CEP of the XUV pulses, and the repetition rate in the XUV pulse train is half the period of the IR pulse. The different contribution from the time-dependent dipole moment factors to the spectra are shown in (c)-(f). The spectrum from the $x_4$ is negligible and not shown. Note that the vertical axis in (d) is in logarithmic scale.

XPA processes are readily contributed by $x_3$. The $x_3$ contribution can be read as the probability for an electron to be ionized by the laser field $E_{IR}(t')$ at time $t'$, propagated from t' to t by the semiclassical action $S_{IR}^{st}(t,t')$, and recombined back to the ground state due to the presence of the attosecond $A_{XUV}(t)$ pulse at time t. Because $d_3(t) \propto A_{xuv}(t)$ [see Eq. (10)], XPA requires signal (seed) from the first gas jet at the same XUV photon energies as the amplified output spectrum, contrarily to the case of enhanced ionization [9-12]. Both the $x_2$ and $x_3$ integrals also include enhanced nonlinear scattering processes by XUV pulses of photon energies far from the ionization potential which are produced in time-durations of the order of few XUV pulse periods and were described in [18]. Figure 3(b)-3(g) shows how the amplifications produced by the $x_2$ and $x_3$ integrals are centered at the same frequency as the seed. The contribution of factor $x_4$ describes a four-wave mixing between the XUV field and the fundamental laser field as l = n ± m ± 1, where l, m and n are harmonic line numbers. It is very small (more than 10-orders of magnitude smaller than $x_1$) and negligible in all cases of our study.

In Figs. 2(c)-2(f) the spectra are obtained from Eq. (7) by computing the complete dipole moment $x(t) \approx x_1(t) + x_2(t) + x_3(t)$ and the separated contributions from $x_1(t)$, $x_2(t)$ and $x_3(t)$, as indicated. As it was commented above, the contribution from $x_4(t)$ is negligible. From the single atom calculations in Figs. 2(c)-2(f) we can therefore observe that the regular HHG spectrum is given by the $x_1(t)$ factor (blue solid line in (c) and (d)), and that the main contribution to the amplification is from the $x_3(t)$ factor (black solid lines in (e) and (f)). The amplification is completed by the contribution from the $x_2(t)$ factor (orange dashed lines in (e) and (f)). The higher XUV amplification obtained in the case of the XUV pulse train [Fig. 2(d) and (f)] compared to the amplification of the single attosecond XUV pulse [Fig. 2(c) and (e)] is obviously due to the larger XUV energy contained in the case of the seed train, since we take a super-Gaussian envelope for the train involving 11 subpulses, each with the same peak intensity as in the single-pulse case ($7\times10^4$ W/cm$^2$), together with the spectral modulation due to the interference fringes of the seed train.

### B. Effect of the delay between the IR and XUV pulses

As it was reported first in Ref. [18] and corroborated experimentally in Ref. [23], the synchronization of the XUV and IR pulses is essential for XUV amplification. We next show by single atom calculations the effect of the delay between the IR pulse and the XUV pulse on the amplification. The single atom calculations are the basis for understanding the more complicated effects included in the propagation of the coherent radiation in the gas medium, which will be treated later both experimentally and theoretically.

We consider a driving laser field composed of a Gaussian temporal profile 26 fs (FWHM) 800 nm IR strong pulse of $7\times10^{14}$ W/cm$^2$ peak intensity, carrier-envelope phase CEP = 0, which produces high-order harmonics in helium ($I_p$ = 24.59 eV) with a photon energy cut-off at $\simeq$ 150 - 160 eV, together with a super-Gaussian 15 fs envelope train of Gaussian XUV 200 as (FWHM) pulses of CEP = 0 and with central photon energy well in the plateau of the IR-generated HHG spectrum. The peak intensity of the XUV attosecond subpulses is only 700 W/cm$^2$ in the present simulations (i.e. $10^{-12}$ times the IR peak intensity), and the temporal separation between the subpulses in the train is half the IR pulse period [see Fig. 2(b)].

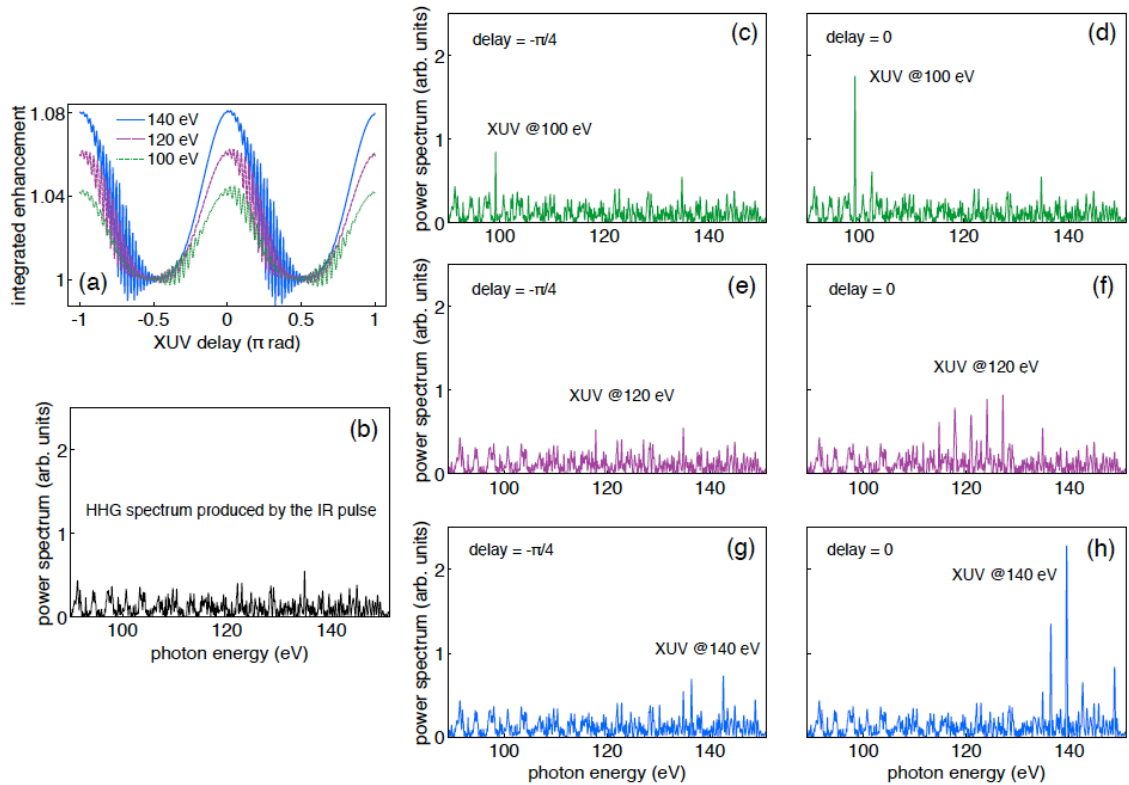

FIG. 3. (Color online). Delay dependent amplification. Single atom calculations for the case of a weak attosecond XUV pulse train interacting together with a strong IR pulse with He ($I_p$=24.587 eV). The IR pulse is a 800 nm, $7\times10^{14}$ W/cm$^2$ peak intensity pulse of 26 fs, CEP=0. The XUV field consists on Gaussian 200 as pulses, CEP=0, with peak intensity of 700 W/cm$^2$ (i.e. $10^{-12}$ times the IR peak intensity). The XUV pulse train has a super Gaussian envelope of 15 fs width. The repetition rate of the train corresponds to half the period of the IR pulse [see Fig. 2 (b)]. (a) Integrated HHG yield enhancement for different values of the central photon energy of the XUV pulse train, as indicated, as a function of the delay between the IR pulse and the XUV pulse train. The delay is given in parts of half a cycle of the IR pulse, in radians, so that $2\pi$ rad $\equiv \lambda_{IR}/c \simeq 2.66$ fs. Note that for the weak XUV peak intensity used in the present calculation the single-atom enhancement is small. (c), (e) and (g) Spectra for the case that the XUV train peak is advanced 0.3325 fs (-$\pi$/4 rad) with respect to the IR pulse. (d), (f) and (h) Spectra for the case that the XUV train peak and the IR pulse peaks are perfectly synchronized. From up to down the XUV pulse trains centered at 100 eV, 120 eV, and 140 eV (green, purple and blue dashed curves, respectively) are shown, and (b) the black solid line show the HHG spectrum obtained with the IR laser pulse alone.

Figure 3(a) shows the amplification obtained by varying the delay of the XUV pulse train with respect to the IR pulse. The simulations in Fig. 3(a) have been performed for different values of the central photon energy of the XUV pulse train, as indicated. The plotted enhancement factor is defined as the integrated HHG yield in the plateau obtained by using the combination of IR+XUV as input pulse divided by the integrated HHG yield obtained considering only the IR pulse. Note that because the integration has been performed over a wide spectral range, the enhancement factor is only somewhat larger than 1. Looking to one harmonic line however the enhancement is much larger. Furthermore, the enhancement in Fig. 3(a) is calculated for a single atom. When one considers propagation in the gas medium the calculation is repeated through several iteration steps and the enhancement becomes large, as it will be shown below.

The fast oscillations in Fig. 3(a) are due to the interference between the harmonics generated by HHG and the amplified XUV train. The period of the fast oscillations coincides with the XUV period, as it was already reported in Ref. [18]. Indeed, the amplified XUV is emitted at the time that the XUV pulse interacts with the medium, so that the fast oscillations are only present where the HHG and amplified XUV fields overlap in time.

Figure 3(b) shows the spectra produced by the interaction of the strong IR pulse alone with the medium (black solid curve) and Fig. 3(c)-3(g) the spectra produced by the interaction of the combination of the strong IR and a weak XUV pulse train centered at 100 eV, 120 eV and 140 eV, as indicated. Clearly, the yield spectra obtained with the combination of IR+XUV input pulses shows an enhancement in the spectral region around 100 eV, 120 eV and 140 eV (green, purple and blue dashed curves, respectively), as it was already shown in the case computed in Fig. 2. Here we show the dependence of this enhancement on the synchronization between the XUV and the IR pulses. In Fig. 3(c), (e) and (g) the XUV train is advanced in time by 0.3325 fs ($\pi/4$ rad) with respect to the peak of the IR pulse field, and this geometry produces a weak amplification of the HHG yield around the corresponding spectral region (100 eV, 120 eV and 140 eV). When the XUV train is synchronized to the peak of the IR field strength, however, which is the case shown in Fig. 3(d), (f) and (h), the amplification is much larger. The dependence of the yield enhancement on the IR field strength is indeed expected from the theory considering the linear dependence of the time-dependent dipole moment factors $x_2(t)$ and $x_3(t)$ on the $E_{IR}$ field [see Eq. (7)].

## IV. AMPLIFICATION OF ATTOSECOND PULSE TRAINS

In this section, we demonstrate that a strong field driven gas behaves as an optical amplifier in the XUV regime by comparing experimental and theoretical results. Both characteristics typical for an optical amplifier are observed, namely the avalanche effect during propagation and the saturation of amplification at high seed fluence.

### A. Saturation of the amplification at increased seed

In general, the output signal of an optical amplifier can be analytically described with the well-known formula [25]

$$j = \ln\left[1 + G_0\left(e^{j_{seed}} - 1\right)\right], \tag{15}$$

where $j = J/J_{sat}$ and $j_{seed} = J_{seed}/J_{sat}$ are the output and the input (seed) fluence, respectively, both are normalized to the saturation fluence; $G_0 = e^{g_0} = e^{\sigma nL}$ is the small signal gain, where σ is the stimulated emission cross-section, $n$ is the atomic density and $L$ is the length of the amplifying medium; and the net gain $G$ can be calculated as $G = j/j_{seed}$. For realizing an efficient amplifier we have to assume $g_0 > 1$ and $G_0 \gg 1$.

Dependent on the seed fluence, one can distinguish between two operational regimes, namely the linear amplifier

$$j_{seed} \ll 1 \,,\ j \approx G_0 j_{seed} \text{ or } J \approx G_0 J_{seed}, \tag{16}$$

and the saturated amplifier

$$j_{seed} > 1 \,,\ j \approx g_0 + j_{seed} \text{ or } J \approx g_0 J_{sat} + J_{seed}. \tag{17}$$

One of the well visible differences between the features of linear and saturated amplifier can be seen on Eq. (16) and (17), namely while in both cases the output is linearly proportional with the seed, the slope in the case of the saturated amplifier is 1 contrary to slope of $G_0$ of the former.

To demonstrate saturated amplification experimentally, the seed XUV fluence was varied by changing the gas pressure (neon) in the seed jet within 50-800 mbar range [see inset of Fig. 4] and the helium pressure in the amplifier jet was hold to 4 bar. The input and output fluences of the amplifier were measured by recording the beam profiles after 200-nm-thick Zr and 200-nm-thick Ti filters. Fig. 4 shows the measured output fluence (red diamonds) in the full scanning range of the seed fluence between $3 \times 10^7$ and $4.7 \times 10^9$ ph/cm². At saturation, the output fluence of the amplifier is mainly determined by the seed fluence, thus the output (XPA) fluence is linearly proportional to the seed fluence as it can be seen on Eq. (17). This linear dependence with an expected slope of 1 is well observable and noted by the fitted (black dashed) line in Fig. 4.

The calculations shown in Fig. 4 are based on the solutions of Eq. (1). The central wavelength of the XUV pulse train is assumed to be 11 nm (central photon energy of $\simeq 113$ eV). Note that the optimal delay between the IR pulse and the central peak of the XUV pulse train is slightly shifted to positive delays in the case of low seed intensity. It can be explained by the interference between the amplified signal generated by the train of pulses and the generated harmonics, and results in this case $\simeq 0.14\ \pi$ rad. At higher values of the seed intensity such interferences become negligible [18] and the optimal delay is basically centered at 0 rad, as it is the case shown in Fig. 3 (a). In Fig, (4) the parameters are the same as used for the calculations summarized in Fig. 3, only the seed intensity is varied and the fluences are scaled in order to fit the measurements. As can be seen in Fig. 4, the results of the numerical calculation (red curve) are in excellent agreement with the measured data.

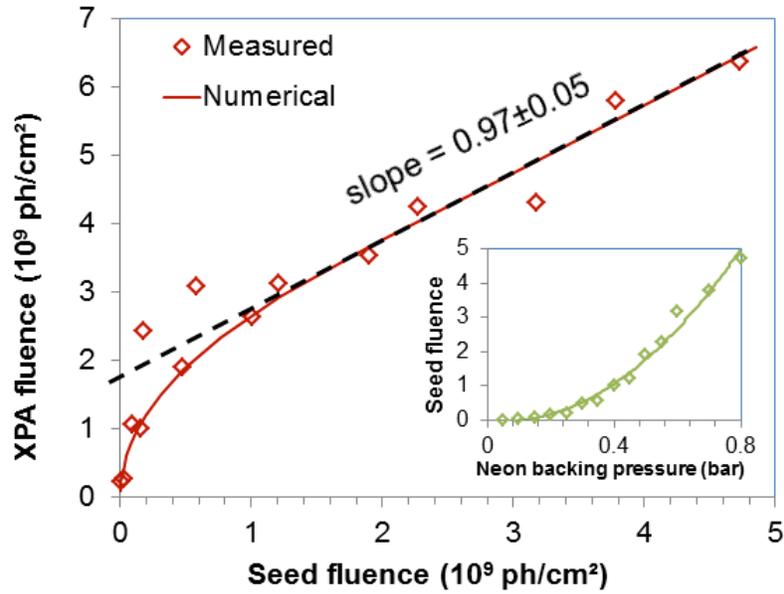

FIG. 4. (Color online) X-ray parametric amplifier behaves as saturated amplifier. The numerical calculation (red line) fits very well to the measured points (red diamonds). At high seed fluence, the output fluence scales linearly with slope $\simeq 1$ in agreement with Eq. (17). (Inset) In the measurement, the seed pulse fluence was varied by changing the applied gas pressure of neon in the seed jet.

## B. Simulation of the XUV pulse propagation in the amplifier jet

At low seed fluence, Eq. (16) describes an avalanche like behavior because the signal increases exponentially along the amplifier medium with length $L$ and density $n$, namely $J_{out} \approx J_{seed} e^{\sigma n L}$. Secondly, the amplifier is linear, because the output fluence is linearly proportional to the seed fluence. For low seed energy, the signal to noise ratio of the measurement was too small to extract reliable information for the linear amplification range directly from the data presented in Fig. 4, consequently we examined this behavior in another way, by changing the atomic density of the amplifier medium to observe the exponential dependence of the gain. The experimental arrangement was the same as presented in Fig. 1. The backing pressure of the neon gas in the seed jet now was fixed at 0.8 bar yielding a seed fluence of $4.7 \times 10^9$ ph/cm², which is in the saturation regime. In the second gas jet, which served as an XUV amplifier, the backing pressure of the helium gas has been varied in a range up to 5 bar. A few measured spectra are plotted in the left column of Fig. 5. We have measured the spectrum of the seed beam (no gas in the amplifier jet, first row) and the spectra of the amplified beams (brown dashed curves) for different settings of the He gas backing pressures in the amplifier jet. Harmonics were also generated in the amplifier jet without the seed beam and we term this case as "unseeded" amplifier. Indeed, the generated harmonics in the amplifier jet are also amplified in the same gas medium, which acts as a self-seeded amplifier. These spectra are also plotted in Fig. 5 (black solid curves). When the amplifier is seeded by an independent external HHG source then we will term this amplifier simply as "seeded" amplifier.

The results from the simulations are shown in the right column of Fig. 5. For the simulations of the experimental measurements, the calculations from Eq. (1) have been adapted to the particular experimental configuration. We used a very simple version of the "particle-in-cell" simulation without calculating any material exchange between the cells. One cell contained ~3800 atoms. The effect of propagation and/or gas pressure was modeled with the increasing number of iteration and we fitted the calculation by set 30 iterations being equal to 5 bar gas pressure. A seed field is first produced by HHG from an intense IR pulse in Ne (first raw). This seed pulse combined at the optimal delay with the intense IR pulse is used as input for the interaction with a first cell of He atoms. The HHG output from this first interaction together with the seed and IR pulses are used as input for a second interaction with a second cell of He atoms, and the process is repeated iteratively, so that propagation is described in 1D and hence we neglect transverse spatial effects. Plasma dispersion has been considered together with neutral dispersion and absorption calculated from the scattering cross sections ($f_1$ and $f_2$) in He, for the propagation of the IR, seed and harmonics fields, with data obtained from Ref. [26]. Indeed, for the propagation of the fields, the accurate neutral dispersion for He together with plasma dispersion are basic ingredients in this study, since group velocity dispersion changes the delay between the different propagating harmonics and the IR field and therefore modifies the overlap between the IR field with respect to the attosecond pulses. Hence propagation effects and the initial delay between seed and IR pulses determine the precise spectral region that is amplified. In this sense, it is important to stress that, as in the experiments, no spectral filter is applied to the HHG output from the first Ne gas jet, which is used as seed pulse for the second He jet by only scaling the value of its yield in order to match the experimental conditions, and therefore no particular spectral region is embedded in the seed pulse for amplification.

Despite the calculations describe the overall scaling of the experimental data very well, some difference between them are obvious, namely for higher pressure the calculated spectra are narrower than the measured ones. This is probably the consequence of assuming a spatially uniform field distribution for the calculations i.e. supposing a plane wave, while in the experiments the profile of the laser beam was near Gaussian and the beam parameters changed somewhat by passing through the gas jet having finite length.

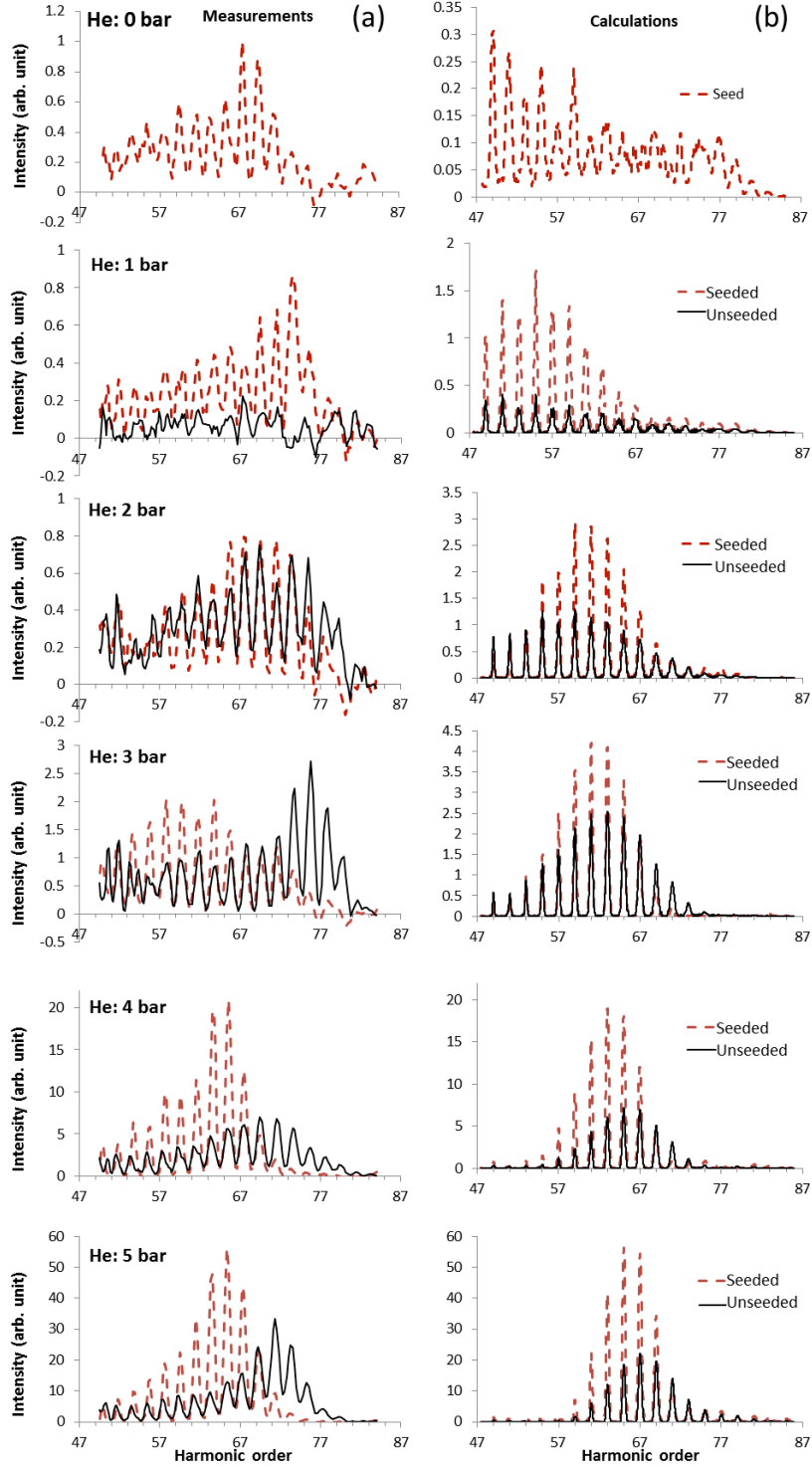

FIG. 5. (Color online) (a) Measured and (b) calculated spectra at increasing gas pressures and corresponding iteration numbers in the second gas jet, respectively. (a) In the He parametric

amplifier, the generated high-harmonic spectra are altered when the amplifier is seeded (brown dashed curves) from an independent HHG source compared to the spontaneously generated spectra without seed (black solid curves). The spectra are normalized to the seed and so directly show the magnitude of the amplification. (b) The numerical simulations reproduce very well the behavior of the spectra of the seeded and unseeded amplifier.

As it is evident by comparing the experimental results with the simulations in Fig. 5, both the measured and calculated spectra show the same behavior. Without applying any seed, there is a continuous increase of the spectral intensity, which also can be seen by the black curves in Fig. 6. However, for the seeded amplifier, the spectra hardly change at low pressures and strong amplification can be observed at higher pressures. To study this behavior in more detail, we plotted the pressure dependence of the spectrally integrated intensity of a few harmonic lines (both measured and calculated) separately in Fig. 6.

For both the seeded (brown dashed) and unseeded (black solid) amplifier, the calculated curves fit very well to the measurement points for harmonics between 63 and 69, where the measured and calculated spectra were most intense. For every harmonics but especially for harmonics 65 and 67, the exponential increase of the harmonic signal in the case of the unseeded amplifier extends over three orders of magnitude. This exponential increase is the clear indication of the avalanche effect of the parametric amplification.

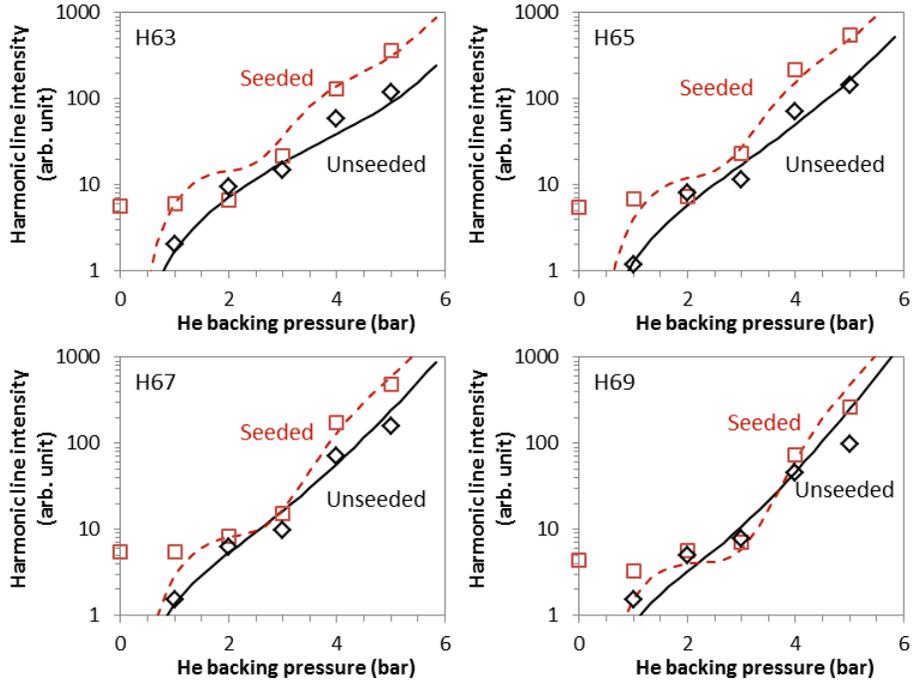

FIG. 6. (Color online) The avalanche effect in HHG can be observed during propagation (or increased He gas pressure) in the amplifier jet. In the different panels, the spectrally integrated intensities of few harmonic lines are plotted. Both the measurements (marks) and the calculations (solid/dashed lines) were performed for the seeded (brown dashed lines) and the unseeded (black solid lines) amplifier.

A closer inspection reveals a more complex behavior for the seeded amplifier which is true for the measurements and the calculations. First, in the calculations, the necessary seed energy is about 10-times smaller to obtain the same harmonic signal as in the experiment. This difference can be clearly seen at low gas pressure (below 1 bar) and supports the assumption that in the measurement probably only about 10% of harmonic beam generated in the first jet was used for seeding the amplifier. This observation is similar than reported in an earlier publication studying XPA at around 300 eV [13]. The difference can be explained by a partial overlap (spatially and temporally) in the amplifier medium. This mismatch is supported by the observation that only a small part of the seed beam was amplified. This observation requires a detailed theoretical study in the future.

**C. Contribution of the parametric amplification and enhanced ionization in HHG**

Another interesting feature of the seeded amplifier is the shoulder between 2 and 3 bar backing pressure. To explain this feature, we performed further calculations by following the $x_1$, $x_2$ and $x_3$ contributions separately during the propagation. The results can be seen on Fig. 7 for the most intense harmonic line of 65. The high harmonic part ($x_1$) remains very small in the full range of propagation or gas pressure. The contribution of enhanced ionization ($x_2$) remains always bellow the XPA, however its rate increases as the XUV signal increases in the medium ($x_3/x_2$, pink). Comparing Fig. 7 with the same H65 of Fig. 6, it is clearly visible that for seeded amplifier the output signal is almost fully governed by the $x_3$ term (XPA) alone and the enhanced ionization and HHG gives only a small contribution. Consequently, both the small value of HHG and the modulation in the XPA are the consequence of the lack of phase matching. We can further observe that the periods and consequently the phase matching conditions for HHG and XPA are somewhat different.

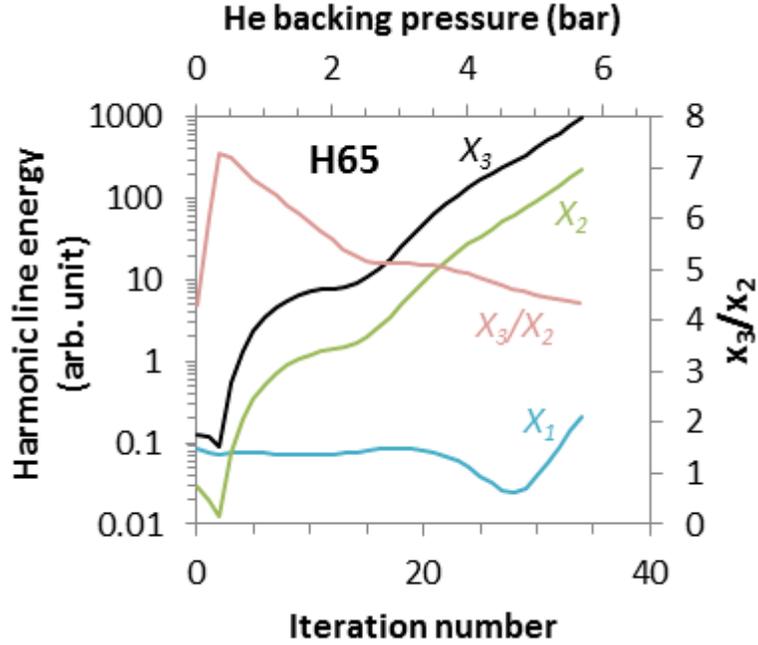

FIG. 7. (Color online) Decomposed calculation of the seeded harmonic source. The three dominating contributions namely the high harmonic generation ($x_1$, blue), the enhanced ionization ($x_2$, green) and the X-ray parametric amplification ($x_3$, black) are plotted separately. The contribution from X-ray parametric amplification dominates during the propagation (iteration number), however its dominance over the effect of the enhanced ionization ($x_3/x_2$, pink) decreases at higher XUV intensities.

## V. DISCUSSION AND CONCLUSIONS

We have experimentally and theoretically studied XUV generation in a two-gas-jet arrangement. We have demonstrated that coherent attosecond XUV pulses can be amplified in He gas in the context of high harmonic generation by carefully adjusting the delay between the intense IR laser pulse and the XUV seed pulses. The numerical simulations show that dispersion in the amplifying medium is a key factor to produce XUV amplification in a specific spectral region. We have investigated in detail two characteristics of an optical amplifier, namely the saturation of the amplification at increased seed and the avalanche-type increase of the generated harmonics during propagation, and have found that numerical simulations based on the SFA fully support and reproduce the experimental measurements and describe the He gas medium as an amplifier of the XUV coherent light pulses. To look into the phenomenon, we distinguished three contributions from the theoretical description of the process, namely high harmonic generation, enhanced ionization and X-ray parametric amplification. We find that X-ray parametric amplification dominates over the other processes and determines the main characteristics of the XUV source. Our results indicate the usability of the SFA theory for the prediction of the optimal conditions and the interpretation of high harmonic generation in two-jet experimental geometries in a broad spectral region considering propagation. The present research hence settles know-how for the generation of intense XUV and X-ray coherent ultrashort light pulses at high repetition rates in typical university laboratories.

## ACKNOWLEDGMENTS

The authors gratefully acknowledge the financial support from the Spanish Ministry of Economy and Competitiveness through "Plan Nacional" (FIS2011-30465-C02-02 and FIS2014-51997-R); from the German DFG grant TR18; from the Japanese "Research Foundation for Opto-Science and Technology"; from the EC's 7th Framework Program (grant 284464, Laserlab Europe HIJ-FSU0019152 and HIJ-FSU001975) and ERC starting grant 258603 NAC. The authors acknowledge the support and contribution of the JETI laser team.

**REFERENCES**

[1] M. Hentschel, R. Kienberger, C. Spielmann, G. A. Reider, N. Milosevic, T. Brabec, P. Corkum, U. Heinzmann, M. Drescher, and F. Krausz, "Attosecond metrology.", Nature **414,** 509-513 (2001).

[2] P. M. Paul, E. S. Toma, P. Breger, G. Mullot, F. Auge, P. Balcou, H. G. Muller, and P. Agostini, "Observation of a Train of Attosecond Pulses from High Harmonic Generation.", Science **292,** 1689-1692 (2001).

[3] E. Seres, J. Seres, and C. Spielmann, "X-ray absorption spectroscopy in the keV range with laser generated high harmonic radiation.", Appl. Phys. Lett. **89,** 181919 (2006).

[4] T. Popmintchev, M. -C. Chen, D. Popmintchev, P. Arpin, S. Brown, S. Ališauskas, G. Andriukaitis, T. Balčiunas, O. D. Mücke, A. Pugzlys, A. Baltuška, B. Shim, S. E. Schrauth, A. Gaeta, C. Hernández-García, L. Plaja, A. Becker, A. Jaron-Becker, M. M. Murnane, and H. C. Kapteyn, "Bright Coherent Ultrahigh Harmonics in the keV X-ray Regime from Mid-Infrared Femtosecond Lasers.", Science **336,** 1287-1291 (2012).

[5] J. Seres, E. Seres, B. Landgraf, B. Ecker, B. Aurand, T. Kuehl, and C. Spielmann, "High-harmonic generation and parametric amplification in the soft X-rays from extended electron trajectories.", Sci. Rep. **4,** 4234 (2014).

[6] J. Seres, E. Seres, B. Landgraf, B. Aurand, T. Kuehl, and C. Spielmann, "Quantum Path Interference and Multiple Electron Scattering in Soft X-Ray High-Order Harmonic Generation.", Photonics **2,** 104-123 (2015).

[7] E. Seres, J. Seres, and C. Spielmann, "Time resolved spectroscopy with femtosecond soft-x-ray pulses." Appl. Phys. A **96,** 43-50 (2009).

[8] E. Seres, and C. Spielmann, "Time-resolved optical pump X-ray absorption probe spectroscopy in the range up to 1 keV with 20 fs resolution.", J. Mod. Opt. **55,** 2643-2651 (2008).

[9] A. Heinrich, W. Kornelis, M. P. Anscombe, C. P. Hauri, P. Schlup, J. Biegert, and U. Keller, "Enhanced VUV-assisted high harmonic generation.", J. Phys. B **39,** S275–S281 (2006).

[10] G. Gademann, F. Kelkensberg, W. Siu, P. Johnsson, K. J. Schafer, M. B. Gaarde, and M. J. J. Vrakking, "Attosecond control of electron-ion recollision in high harmonic generation.", New J. Phys. **13,** 033002 (2011).

[11] F. Brizuela, C. M. Heyl, P. Rudawski, D. Kroon, L. Rading, J. M. Dahlstrom, J. Mauritsson, P. Johnsson, C. L. Arnold and A. L'Huillier, "Efficient high-order harmonic generation boosted by below-threshold harmonics." Sci. Rep. **3**, 1410 (2013).

[12] E. J. Takahashi, T. Kanai, K. L. Ishikawa, Y. Nabekawa, and K. Midorikawa, "Dramatic Enhancement of High-Order Harmonic Generation.", Phys. Rev. Lett. **99,** 053904 (2007).

[13] J. Seres, E. Seres, D. Hochhaus, B. Ecker, D. Zimmer, V. Bagnoud, T. Kuehl, and C. Spielmann, "Laser-driven amplification of soft X-rays by parametric stimulated emission in neutral gases.", Nat. Phys. **6**, 455-461 (2010).

[14] A. Fleischer, and N. Moiseyev, "Amplification of high-order harmonics using weak perturbative high-frequency radiation.", Phys. Rev. A **77,** 010102R (2008).

[15] A. A. Svidzinsky, L. Yuan, and M. O. Scully, "Quantum Amplification by Superradiant Emission of Radiation.", Phys. Rev. X **3,** 041001 (2013).

[16] C. Serrat, "Broadband spectral amplitude control in high-order harmonic generation.", Appl. Sci. **2,** 816–830 (2012).

[17] C. Serrat, "Broadband spectral-phase control in high-order harmonic generation.", Phys. Rev. A **87,** 013825 (2013).

[18] C. Serrat, "Coherent Extreme Ultraviolet Light Amplification by Strong-Field-Enhanced Forward Scattering.", Phys. Rev. Lett. **111**, 133902 (2013).

[19] C. Serrat, D. Roca, and J. Seres, "Coherent amplification of attosecond light pulses in the water-window spectral region.", Opt. Express **23**, 4867-4872 (2015).

[20] J. Seres, E. Seres, and C. Spielmann, "Classical model of strong-field parametric amplification of soft x rays.", Phys. Rev. A **86,** 013822 (2012).

[21] L. V. Dao, K. B. Dinh, P. Hannaford, "Perturbative optical parametric amplification in the extreme ultraviolet.", Nat. Commun. **6**, 7175 (2015).

[22] M. Lewenstein, P. Balcou, M. Y. Ivanov, A. L'Huillier, and P. B. Corkum, "Theory of high-harmonic generation by low-frequency laser fields.", Phys. Rev. A **49,** 2117-2132 (1994).

[23] J. Seres, E. Seres, B. Landgraf, B. Ecker, B. Aurand, A. Hoffmann, G. Winkler, S. Namba, T. Kuehl, and C. Spielmann, "Parametric amplification of attosecond pulse trains at 11 nm.", Sci. Rep. **4**, 4254 (2014).

[24] M. V. Ammosov, N. B. Delone, V. P. Krainov, "Tunnel ionization of complex atoms and of atomic ions in an alternating electromagnetic field.", Sov. Phys. JETP **64**, 1191-1194 (1986).

[25] B. C. Stuart, S. Herman, and M. D. Perry, "Chirped-Pulse Amplification in Ti:Sapphire Beyond 1 μm.", IEEE J. Quant. Electr. **31**, 528-538 (1995).

[26] B. L. Henke, E. M. Gullikson, and J. C. Davis, "X-ray interactions: photoabsorption, scattering, transmission, and reflection at E=50-30000 eV, Z=1-92.", Atomic Data and Nuclear Data Tables **54,** 181-342 (1993).